\documentclass[prb,twocolumn,amsmath,amssymb,aps]{revtex4}

\usepackage{amssymb} % Package for more symbols (\rightleftarrows).
\usepackage{graphicx} % Required for the inclusion of images
\usepackage{hyperref} % Hyperlink reference package.

\graphicspath{{./Figures/}} % Specifies the directory where pictures are stored

\begin{document}

\title{Dynamics of quasiparticle trapping in Andreev levels}

\author{D. G. Olivares}
\affiliation{Departamento de F\'{i}sica Te\'orica de la Materia Condensada, Condensed
Matter Physics Center (IFIMAC),
and Insitituto Nicol\'as Cabrera, Universidad Aut\'{o}noma de Madrid, E-28049 Madrid, Spain.}
\author{L. Bretheau}
\affiliation{Quantronics Group, Service de Physique de l'Etat Condens\'{e} (CNRS, URA 2464), IRAMIS, CEA-Saclay, 91191 Gif-sur-Yvette, France.}
\author{\c{C}.~\"O. Girit}
\affiliation{Quantronics Group, Service de Physique de l'Etat Condens\'{e} (CNRS, URA 2464), IRAMIS, CEA-Saclay, 91191 Gif-sur-Yvette, France.}
\author{H. Pothier}
\affiliation{Quantronics Group, Service de Physique de l'Etat Condens\'{e} (CNRS, URA 2464), IRAMIS, CEA-Saclay, 91191 Gif-sur-Yvette, France.}
\author{C. Urbina}
\affiliation{Quantronics Group, Service de Physique de l'Etat Condens\'{e} (CNRS, URA 2464), IRAMIS, CEA-Saclay, 91191 Gif-sur-Yvette, France.}
\author{A. Levy Yeyati}
\affiliation{Departamento de F\'{i}sica Te\'orica de la Materia Condensada, Condensed
Matter Physics Center (IFIMAC),
and Insitituto Nicol\'as Cabrera, Universidad Aut\'{o}noma de Madrid, E-28049 Madrid, Spain.}

\date{\today}

\begin{abstract}
We present a theory describing the trapping of a quasiparticle in a prototypical Josephson junction, a single-channel superconducting weak link.  We calculate the trapping and untrapping rates associated to absorption and emission of both photons and phonons. We show that the presence of an electromagnetic mode with frequency smaller than the gap gives rise to a rather abrupt transition between a fast relaxation regime dominated by coupling to photons and a slow relaxation regime dominated by coupling to phonons. This conclusion is illustrated by the analysis of a recent experiment \cite{longlived} measuring the dynamics of quasiparticle trapping in a superconducting atomic contact coupled to a Josephson junction. With realistic parameters the theory provides a semi-quantitative description of the experimental results.
\end{abstract}

\maketitle

\section{Introduction} 

There are several external mechanisms that
undermine the quantum coherence of superconducting circuits being
explored for quantum information processing \cite{devoret2013}. Their
influence has been reduced over the years by new designs that minimize
the coupling with external degrees of freedom. However, a fundamental
intrinsic decoherence process arises from the coupling of the qubit
variables to superconducting quasiparticles tunneling through the
Josephson junctions of the circuits. Although
in principle the superconducting gap $\Delta$ provides an inherent
protection against low energy excitations at low temperatures, in
practice there are residual nonequilibrium quasiparticles that can
rule the behavior of the circuits \cite{martinis1,martinis2,martinis3,catelani,riste,Geerlings,spectro}. As shown in a recent experiment \cite{Levenson}, this is particularly true for weak links containing channels of high transmission, where localized excitations occupying Andreev levels of energy below $\Delta$ become possible. This has important consequences for the corresponding proposed qubits designs \cite{marce,shumeiko,nazarov,Padurariu}. 
Furthermore, single quasiparticle trapping in localized levels could
be detrimental in experiments proposed to detect ``Majorana bound
states'' in condensed matter systems since their
topological protection relies on parity conservation \cite{majorana}. Understanding
the dynamics of relaxation of quasiparticles in superconducting
weak links is therefore an important present-day issue.

We report here on a theory highlighting the role of the electromagnetic environment in this dynamics. In experiments, Josephson junctions or weak links are very often embedded in electrical circuits having electromagnetic modes at frequencies lower than the superconducting gap. The environment can be a resonator intentionally coupled to the junction like 
in Ref.~\onlinecite{Levenson}, or the plasma mode of another junction placed in parallel like in Ref.~\onlinecite{longlived}. We show that if the mode impedance is large enough, it rules the quasiparticle dynamics when the sum of the  Andreev level energy and of the energy of the mode exceed the superconducting gap.

The rest of the paper is organized as follows: In Sec. \ref{model} we describe the model
considered for a superconducting one channel contact coupled to a generic
electromagnetic environment; Sec. \ref{transitionrates} is devoted to the analysis of the transition
rates between different quasiparticles states induced by quantum phase fluctuations;
in Sec. \ref{comparison} we focus on the experimental situation of 
Ref.~\onlinecite{longlived} and
compare the theoretical results for the transition rates and the stationary probability
for quasiparticles trapped in the subgap states with the corresponding experimental
results. In Sec. \ref{conclusions} we present our main conclusions. The more technical
details on our calculations are described in appendices \ref{appendixA}, \ref{appendixB}, \ref{appendixC} and \ref{appendixD}.  

\begin{figure}[h!]
\begin{centering}
\includegraphics[scale=0.4]{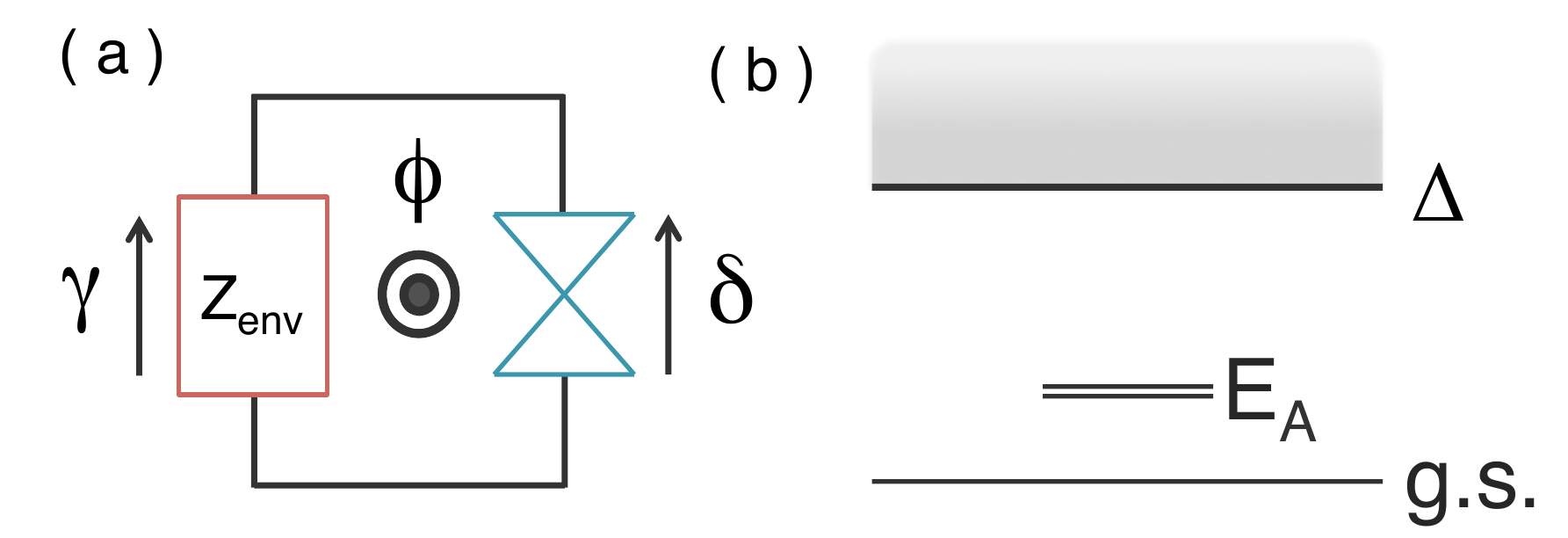}
\caption{(a) Schematic representation of a single superconducting channel coupled to an impedance $Z_{\text{env}}.$  $\delta$
and $\gamma$ indicate the phase drops through the channel and the impedance, respectively, and $\Phi$ is the magnetic flux
through the loop. (b) Quasiparticle excitation spectrum in the single channel weak link, with
the continuum above the gap $\Delta$ and a subgap discrete spin-degenerate Andreev
level of energy $E_A$.}
\label{setup} 
\end{centering}
\end{figure}

\section{Theoretical framework} 
\label{model}
We consider the situation illustrated
in Fig.~\ref{setup}(a) with a single superconducting channel  (SC) coupled to an arbitrary impedance  $Z_{\text{env}}.$
The excitation spectrum of the SC contains a discrete, spin-degenerate Andreev level, with an energy $E_{A}(\delta)=\Delta\sqrt{1-\tau\sin^{2}\delta/2}$,
where $\delta$ is the superconducting phase difference across the
contact and $\tau$ the transmission probability for electrons \cite{beenakker}
(see Fig.~\ref{setup}(b)). The Andreev level is completely empty
when the channel is in its ground state, which has a phase dependent
energy $-E_{A}$ and carries a supercurrent $I=-\left(\partial E_{A}/\partial\delta\right)/\varphi_{0},$ where $\varphi_{0}=\hbar/2e$ is the reduced flux quantum. The 
lowest-energy excitations correspond to the occupation of the Andreev level
by a single quasiparticle (of either spin), the global energy and
the supercurrent of these ``odd" configurations
being then zero. There is also an excitation of energy $2E_{A}$ with 
respect to the ground state, 
where the Andreev level is occupied by two quasiparticles of opposite
spins. This ``even" configuration can be seen as
a localized excited ``Andreev pair" \cite{nature-landry},
and carries a supercurrent opposite to that in the ground state.

The system Hamiltonian can be written
as $\hat{H}=\hat{H}_{SC}(\hat{\delta})+\hat{H}_{\text{env}}(\hat{\gamma})$,
where the first term describes the SC and the second one the electromagnetic environment, modelled by the impedance $Z_{\text{env}}.$ The phases $\delta$ and $\gamma$
across the SC and the impedance are related by $\hat{\delta}-\hat{\gamma}=\Phi/\varphi_{0}=\varphi$,
where $\Phi$ is the magnetic flux through the loop.

The population of the SC electronic states becomes then sensitive
to the effects of quantum phase fluctuations. Assuming that $\text{Re}(Z_{\text{env}}) \ll R_{Q},$
we treat the fluctuations to lowest order in perturbation and write
the Hamiltonian $\hat{H}=\hat{H}_{\text{env}}(\hat{\gamma})+\hat{H}_{SC}(\varphi)+\varphi_{0}\hat{\gamma}\hat{I}(\varphi)$, 
where $\hat{I}= \varphi_{0}^{-1} \partial \hat{H}_{SC}/\partial\delta$
is the current operator in the contact region.

To describe the unperturbed single-channel SC we use a one-dimensional
SNS junction model with a Dirac delta potential barrier (to account
for non-perfect transmission) inside a normal region of negligible length.
Details of the diagonalization of this model in terms of Bogoliubov
fermion operators $\gamma_{\alpha,\sigma}$, where $\sigma$ indicates
spin, are given in Appendix \ref{appendixA}. Two types
of states are obtained, $\alpha\equiv k$ with energy $E_k\geq\Delta$
corresponding to the extended continuum states and $\alpha\equiv A$
corresponding to the localized Andreev states with energy $E_{A}$
(see Fig.~\ref{setup}~(b)). The SC ground state $\left\lvert \Psi_{0}\right\rangle $
corresponds to the absence of excitations, i.e. $\gamma_{\alpha,\sigma}\left\lvert \Psi_{0}\right\rangle =0$.

\section{Transition rates} 
\label{transitionrates}

\begin{figure}
\begin{centering}
\includegraphics[scale=0.35]{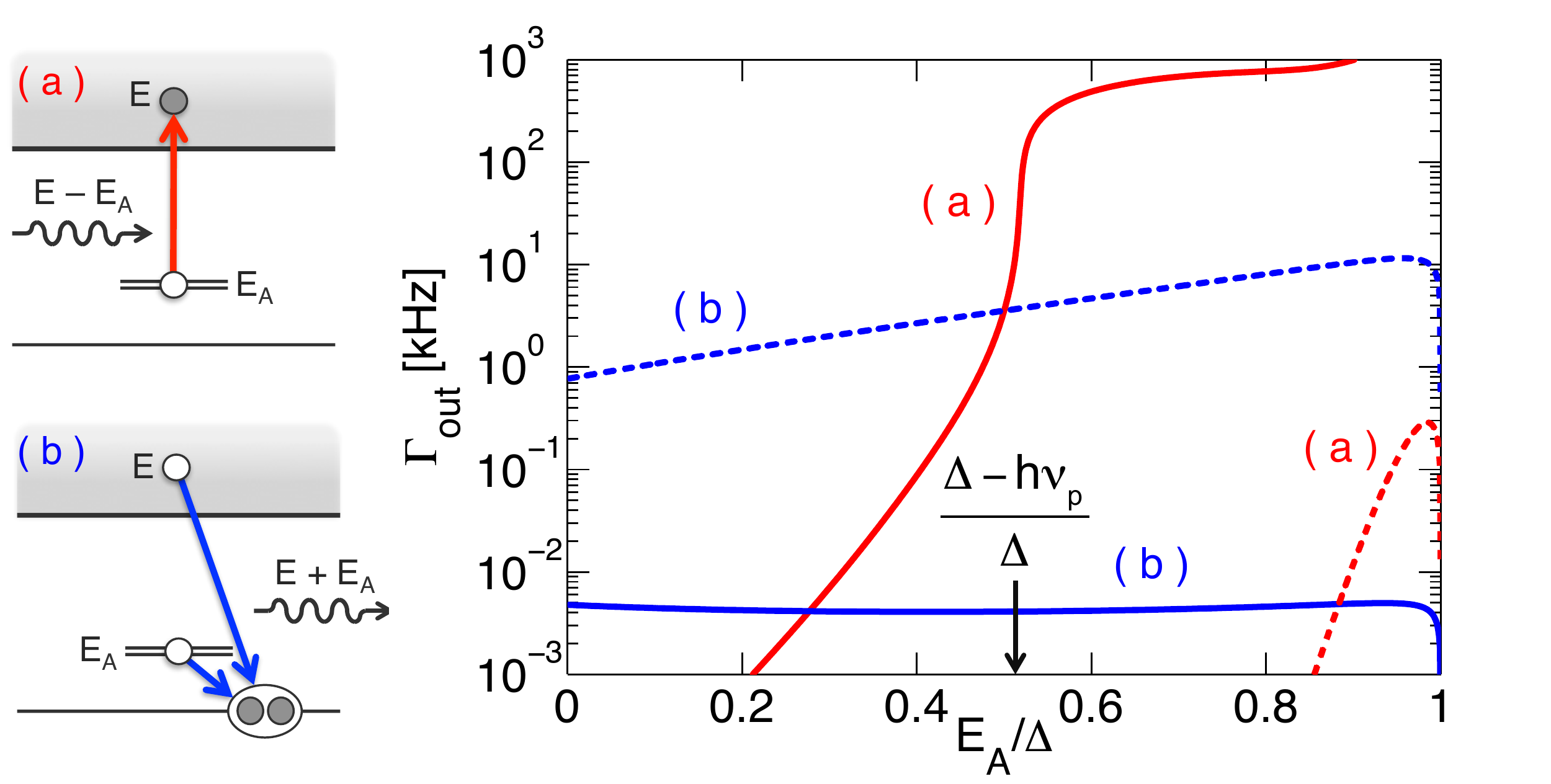}
\caption{Left panel: Schematic representation of processes removing a quasiparticle from
the Andreev level. Process (a) involves absorption of an environmental
photon or phonon and corresponds to the transition rate $\Gamma_{\text{out}}^{(a)}$.
Process (b) corresponds to the recombination into a Cooper pair of the quasiparticle
trapped in the Andreev level with a quasiparticle from the continuum,
with the emission of a photon or a phonon, and
is characterized by a transition rate $\Gamma_{\text{out}}^{(b)}$. The right
panel shows the rates $\Gamma_{\text{out}}^{(a)}$ (red) and $\Gamma_{\text{out}}^{(b)}$
(blue) resulting from the absorption and emission of environmental
photons (full lines) or phonons (dashed lines), for the parameters of Ref.~\onlinecite{longlived} (see Appendix \ref{appendixC}). We have set $k_BT_{\text{env}}=0.06 \Delta$,
$k_B T_{\text{qp}}=0.09 \Delta$, $k_B T_{ph}=0.015 \Delta$.}
\label{results-gamma-out} 
\end{centering}
\end{figure}

The coupling of the SC to the environment
allows for transitions between different quasiparticle states. We
shall first consider processes which permit the removal of a quasiparticle
from the Andreev level. These processes allow in particular the relaxation
of the lowest-energy excited states with one trapped quasiparticle
back to the ground state $\left\lvert \Psi_{0}\right\rangle $ \cite{comment}. They
consist either in the absorption of an environmental photon and transfer
of the trapped quasiparticle into the continuum states, or in the recombination
of a quasiparticle from the continuum with the trapped one into a
Cooper pair while releasing the energy as a photon. These two processes
are illustrated in panels (a) and (b) of Fig.~\ref{results-gamma-out}
and the corresponding rates are denoted by $\Gamma_{\text{out}}^{(a,b)}.$ The Fermi golden rule for the
first process yields
\begin{eqnarray}
\Gamma_{\text{out}}^{(a)} &=& \frac{2\pi}{\hbar} \sum_k  \left|\left\langle k,\sigma\left|\varphi_{0}\hat{I}\right|A,\sigma\right\rangle \right|^{2}P\left(E_k-E_{A}\left(\delta\right)\right)\nonumber\\
&& \times \left( 1 - f_{\text{FD}}(E_k,T_{\text{qp}})\right) \;,
\label{golden-rule}
\end{eqnarray}
where $f_{\text{FD}}\left(E,T_{\text{qp}}\right)$
is the Fermi population factor for quasiparticles in the continuum
(assumed to be in equilibrium at a temperature $T_{\text{qp}}$) and $P(E)$
is the probability of absorbing a photon of energy $E$ from the environment.
This probability is $P\left(E\right)=D\left(E\right)f_{\text{BE}}\left(E,T_{\text{env}}\right)$,
where $f_{\text{BE}}\left(E,T\right)$ is the Bose population factor, and
$D\left(E\right)=\text{Re}\left\lbrace Z_{\text{env}}\left(E\right)/E\right\rbrace /R_{Q}$,
with $R_{Q}=h/4e^{2}$, is the density of states for the modes in
the environment \cite{IngoldNazarov}. The environment is assumed
to be in equilibrium at a temperature $T_{\text{env}}$ which can be in general
different from $T_{\text{qp}}$. The numerical evaluation of this rate (and every
other) for different transmissions, shows a rather universal
dependence in the Andreev level energy position $E_{A}$. Simple analytical expressions
can be derived in the perfect transmission limit $\tau\rightarrow1$
and in the tunnel limit $\tau\rightarrow0$,
for which the wavefunctions have a considerably simpler form. In this
limit one obtains (see Appendix \ref{appendixB})
\begin{eqnarray}
\Gamma_{\text{out}}^{\left(a\right)} &=&  \frac{8\Delta}{h}\int_{\Delta}^{\infty} dE D\left(E-E_{A}\right)g\left(E,E_{A}\right)  \nonumber \\
&& \times  f_{\text{BE}}\left(E-E_{A},T_{\text{env}}\right) \left(1-f_{\text{FD}}\left(E,T_{\text{qp}}\right)\right),
\label{gamma-out-1}
\end{eqnarray}
with %$g\left(E,E_{A}\right)=\frac{\sqrt{E^{2}-\Delta^{2}}\sqrt{\Delta^{2}-E_{A}^{2}}}{\Delta\left(E-E_{A}\right)}.$
$g\left(E,E_{A}\right)=\sqrt{\left(E^{2}-\Delta^{2}\right) \left(\Delta^{2}-E_{A}^{2}\right)} / (\Delta\left(E-E_{A}\right)).$
We also give in Appendix \ref{appendixB} the expression of $g\left(E,E_{A}\right)$ in the tunnel limit $\tau \rightarrow 0.$

When the environment of the SC contains a single mode with infinite quality factor, and at low temperature, this expression simplifies to $\Gamma_{\text{out}}^{\left(a\right)} = \frac{2\Delta}{\hbar} \frac{Z_{0}}{R_{Q}} g\left(E_{A}+h \nu,E_{A}\right)\text{exp}\left(-h \nu /k_{B}T_{\text{env}}\right),$ where $\nu$ is the mode frequency. The function $g$ is of order 1 when $ \Delta-h \nu  < E_{A} <  \Delta, $ so that the rate is simply determined by the impedance  $Z_{0}$ of the oscillator. For aluminum and for $Z_{0}=50~ \Omega$, $\frac{2\Delta}{\hbar} \frac{Z_{0}}{R_{Q}} \approx 1~ \text{GHz.}$

In a similar way for the second relevant process we find 
\begin{eqnarray}
\Gamma_{\text{out}}^{\left(b\right)} &=& \frac{8\Delta}{h}\int_{\Delta}^{\infty} dE
D\left(E+E_{A}\right)
g\left(E,-E_{A}\right)
\nonumber \\
&& \times \left(1+f_{\text{BE}}\left(E+E_{A},T_{\text{env}}\right)\right) 
f_{\text{FD}}\left(E,T_{\text{qp}}\right).
\label{gamma-out-2}
\end{eqnarray}

We show in Appendix \ref{appendixD} that for perfect transmission
the matrix elements for electron-phonon coupling have the same functional
form in terms of $E$ and $E_{A}$ as those for the coupling with
the electromagnetic modes. 
Therefore its inclusion leads to the same
expressions for $\Gamma_{\text{out}}^{(a,b)}$ as in Eqs. (\ref{gamma-out-1},\ref{gamma-out-2})
but with a quadratic density of states and $T_{\text{env}}$ replaced by the phonon temperature $T_{ph}$ (see Appendix \ref{appendixD} for more
details). 

\begin{figure}
\begin{centering}
\includegraphics[scale=0.35]{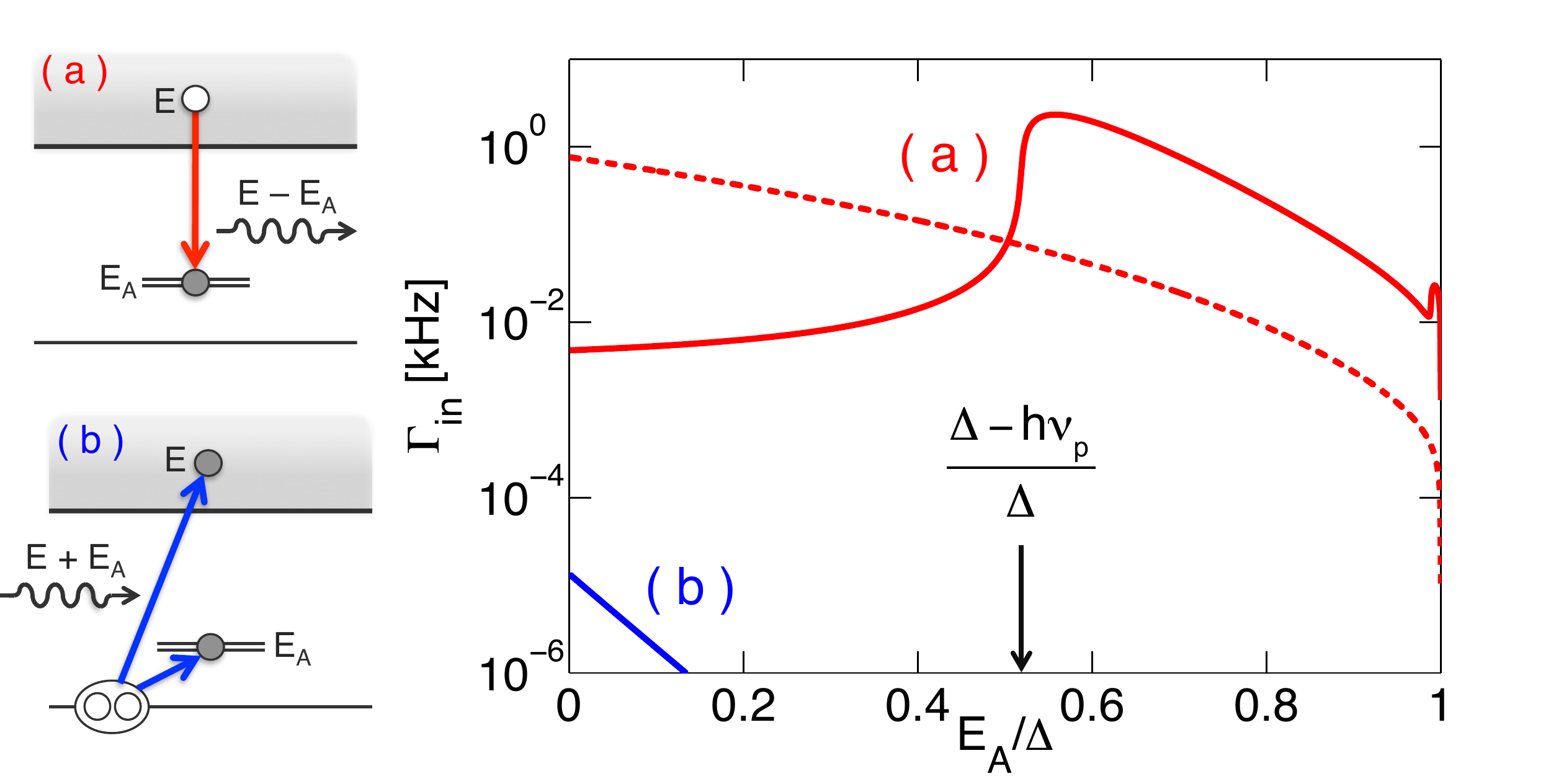}
\caption{Left: Schematic representation of processes adding a quasiparticle
in the Andreev level. Right: the emission processes (a) rates, denoted
by $\Gamma_{\text{in}}^{(a)}$ in the text, are given by the red curves for
photons (full line) and phonons (dashed lines). The processes (b)
involving the breaking of a Cooper pair (blue line in the plot)
are much less efficient. The one involving the phonons is below this scale. Same parameters
as in Fig. \ref{results-gamma-out}.}
\label{results-gamma-in}
\end{centering}
\end{figure}

The time reversed processes, illustrated in Fig.~\ref{results-gamma-in}
and characterized by rates $\Gamma_{\text{in}}^{(a)}$ and $\Gamma_{\text{in}}^{(b)}$
are responsible for the population of the Andreev level, either by
trapping a quasiparticle from the continuum or by breaking a 
pair.

\section{Comparison to experiments} 
\label{comparison}

We focus on the recent experiments on superconducting atomic contacts \cite{longlived}
that have analyzed in detail the quasiparticle trapping in Andreev levels and its dynamics. In these experiments, an atomic contact was embedded in a superconducting
loop containing a Josephson junction, thus forming an asymmetric SQUID. It was found that there is a significant
probability for the SC to get trapped in an odd state in which the
highest transmitted channel carries no supercurrent. The experiments
also showed that the relaxation rates for these states fall into a
nearly universal behavior as a function of the energy $E_{A}$ regardless
of the particular values of the transmission and phase difference.
Trapping occurred essentially when the Andreev level energy was smaller
than half the superconducting gap $E_{A}\lesssim0.5\Delta$, with
the lifetime of trapped quasiparticles exceeding 100~$\mu$s. For
larger energies no significant trapping could be detected. The origin
of this sharp energy threshold was a puzzle not explained in the paper reporting
the experiment.

In Ref.~\onlinecite{longlived}, the SQUID  Josephson junction had a Josephson energy much larger than the charging energy, and it can therefore be described as an harmonic oscillator. Spectroscopy measurements \cite{nature-landry,landrythesis} on similar circuits as the one used in Ref.~\onlinecite{longlived} showed that the plasma frequency of this mode can be significantly renormalized by parallel inductances and approach
$\sim0.5\Delta$ (see Appendix \ref{appendixC} and Ref.~\onlinecite{landrythesis}). As explained in the following, our theory shows that the main relaxation mechanism for the trapped quasiparticle states is their excitation into the extended continuum states above the superconducting gap by absorption of photons from the plasma mode. This mechanism becomes inefficient when the energy difference between the Andreev level and the continuum exceeds the plasma energy, $\Delta - E_A > h\nu_{p},$ hence providing a simple explanation for the observed behavior.

The results for the transition rates $\Gamma_{\text{out}}^{(a,b)}$ obtained using parameters which are appropriate for the experimental situation
of Ref.~\onlinecite{longlived} (see Appendix \ref{appendixC})
are shown by the solid lines in Fig.~\ref{results-gamma-out}.
For $E_{A}\gtrsim\Delta-h\nu_{p}\sim0.52\Delta$, $\Gamma_{\text{out}}^{(a)}$
is large because photons in the plasma
mode can excite the trapped quasiparticle out into the continuum. Similarly, $\Gamma_{\text{in}}^{(a)}$
is large in this energy range because quasiparticles near the gap edge can relax in the Andreev level while emitting a plasma photon (see Fig.~3). 
For lower energies, the energy of the plasma photons is not sufficient
and the rate drops abruptly.
Other processes, like phonon absorption or emission start to play
a role. Hence,
both $\Gamma_{\text{in}}$ and $\Gamma_{\text{out}}$ are determined by phonon processes
for $E_{A}<\Delta-h\nu_{p}$ and by photon processes for $E_{A}>\Delta-h\nu_{p}$. It should be noticed that three different temperatures enter the 
calculation. We assume that the phonons in the Al films of Ref.~\onlinecite{longlived} are at equilibrium with the substrate and therefore $T_{ph}$ is taken equal to the base temperature measured by the thermometers 
in the experiment ($30\,\text{mK}$). The two other temperatures, $T_{\text{env}}$ and $T_{\text{qp}}$, can be significantly larger due to incomplete filtering of radiation. 
To fit the results we have used $T_{\text{env}} \sim 120 \,\text{mK}$, similar to what is deduced from measurements of the switching probability of the SQUID \cite{supplemental-longlived,landrythesis}
and $T_{\text{qp}} \sim 180\,\text{mK}$ which simulates the presence of a few tens of out-of-equilibrium quasiparticles per $\mu\text{m}^3$, as typically found in experiments 
with Al resonators and qubits \cite{martinis1,martinis2,martinis3,riste,Klapwijk}. 

\begin{figure}
\begin{centering}
\includegraphics[scale=0.35]{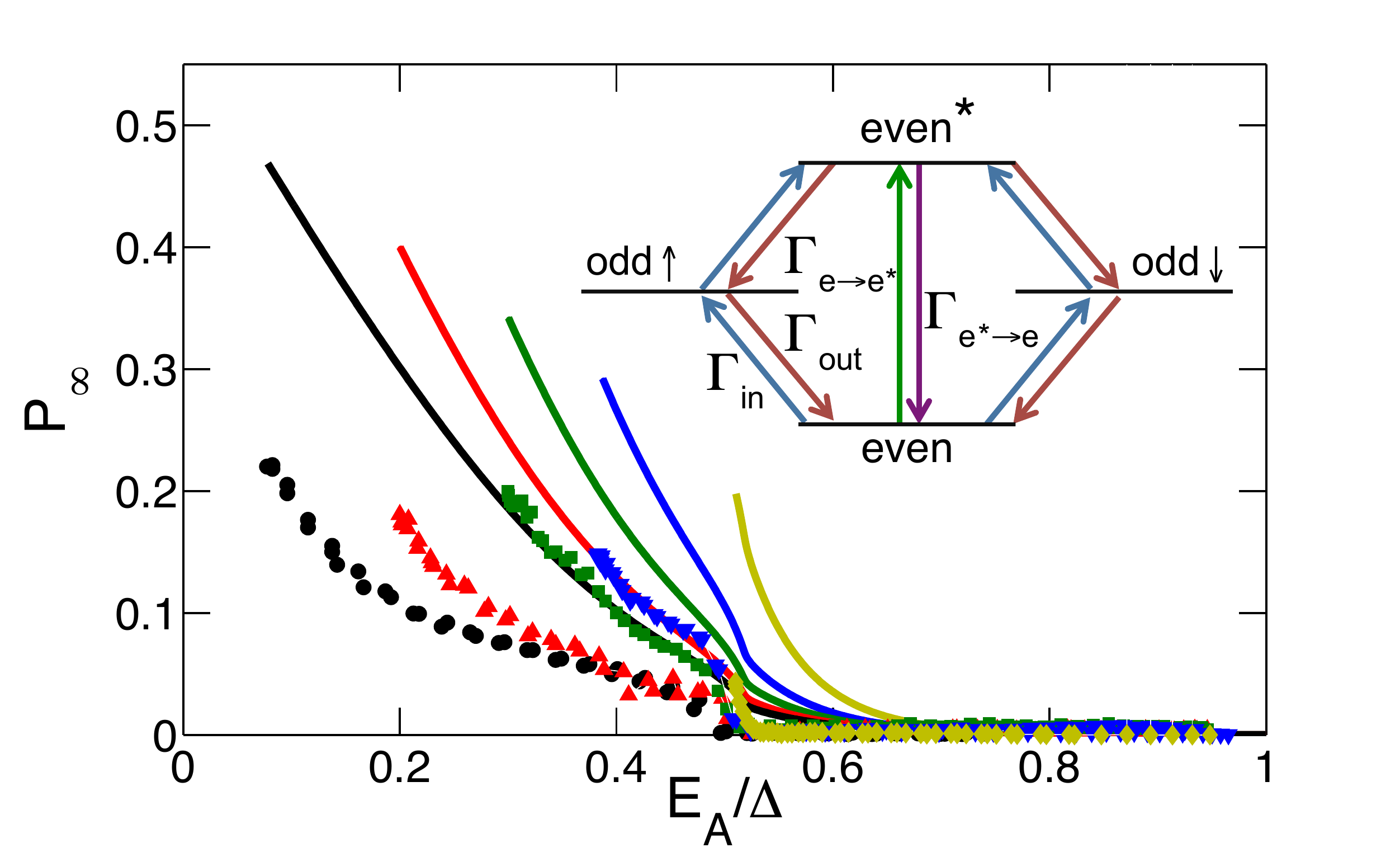}
\caption{Stationary occupation probability $P_{\infty}$ for the odd states, using parameters of Ref.~\onlinecite{longlived}.
The theoretical results (full lines) are compared with the experimental
results for different values
of the contact transmission $\tau=$ 0.994 (full circles, black);
0.96 (upper triangles, red); 0.91 (squares, dark green); 0.85 (down
triangles, blue) and 0.74 (diamonds, light green).
Same parameters as in Figs.~\ref{results-gamma-out} and \ref{results-gamma-in}.
(Inset) Scheme of the Andreev level occupation configurations and
the different transitions induced by the coupling to the environment.
The total rates $\Gamma_{out,in}$ connecting odd states with the
even states (red and blue arrows) are obtained by adding $\Gamma_{out,in}^{(a)}$ and
$\Gamma_{out,in}^{(b)}$. The rates $\Gamma_{e^{*}\rightarrow e}$
and $\Gamma_{e^{*}\rightarrow e}$ connecting the even states (green and purple arrows)
are calculated within the same
model (see Appendix \ref{appendixB}).}
\label{p-infinity} 
\end{centering}
\end{figure}

The different transitions which determine the population and relaxation
of the Andreev level are illustrated in the inset of Fig. \ref{p-infinity}.
They also involve the 
even excited state $\left\lvert \mbox{even}^{*}\right\rangle =
\gamma_{A \uparrow}^{\dagger}\gamma_{A \downarrow}^{\dagger}\left\lvert \Psi_{0}\right\rangle$.
The analysis is further simplified by symmetry relations:
the rates connecting the
even excited state and the odd states are equal to the ones connecting
the odd states and the even ground state. This is indicated by the
color code used for the arrows in the inset of Fig. \ref{p-infinity}.
Notice that the full determination of the level populations requires
also the evaluation of the rates $\Gamma_{e^{*}\rightarrow e}$ and
$\Gamma_{e\rightarrow e^{*}}$. In Ref.~\onlinecite{longlived} it was assumed
that the relaxation rate $\Gamma_{e^{*}\rightarrow e}$ to the ground
state was very fast compared to all other ones and that $\Gamma_{e\rightarrow e^*}$
was negligible. For photonic or phononic
environments these have been calculated in Refs.~\onlinecite{marce,landrythesis}
and \onlinecite{zazunov2005} respectively and reproduced with the present
formalism, as discussed in Appendix \ref{appendixB} and \ref{appendixD}. The calculation does corroborate that $\Gamma_{e^{*}\rightarrow e}$
is much larger than all the other rates for the transmissions explored
in the experiment. 

A last step in our calculation is to obtain the stationary distribution of quasiparticles by solving the master equation
involving all transitions indicated in the inset of Fig.~\ref{p-infinity}.
The result for the occupation probability of the odd states, $P_{\infty}$,
is shown in Fig.~\ref{p-infinity} and compared with the experimental
results from Ref.~\onlinecite{longlived} for contacts with different transmissions.
As can be observed, the theory qualitatively
describes the decrease in $P_{\infty}$ at fixed $E_{A}$ which is
observed experimentally for increasing transmission in the slow relaxation
regime. There is, however, some discrepancy in the quantitative values
of $P_{\infty}$ which is overestimated in our model calculations.

\begin{figure}
\begin{centering}
\includegraphics[scale=0.35]{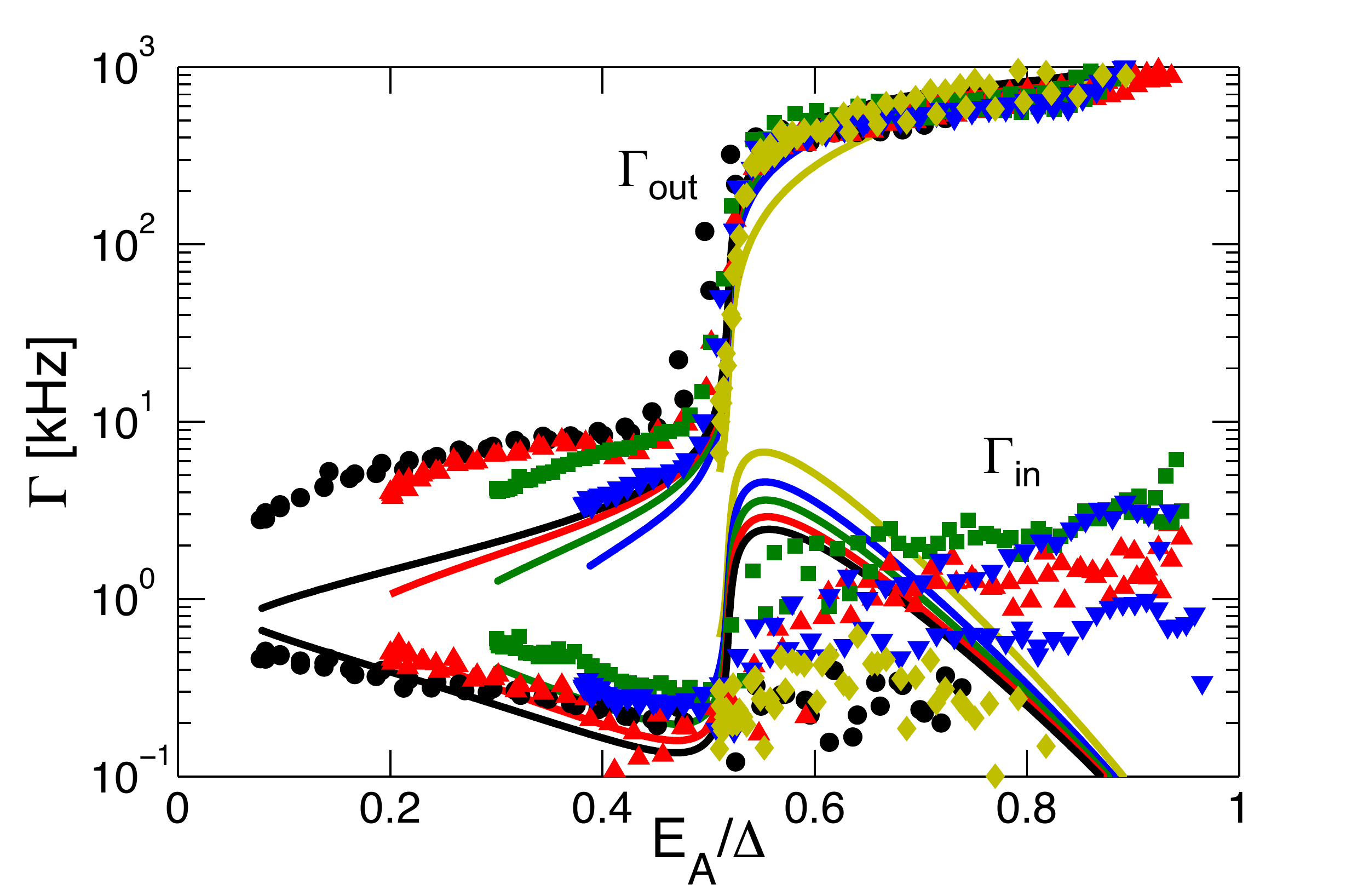}
\caption{Total rates $\Gamma_{\text{out}}$ (upper curves and symbols) and $\Gamma_{\text{in}}$
(lower cuves and symbols) calculated by the present model and obtained
experimentally in Ref.~\onlinecite{longlived} for different contact transmissions.
The same convention as in Fig.~\ref{p-infinity} is used.}
\label{full-rates-results} 
\end{centering}
\end{figure}

We show in Fig.~\ref{full-rates-results} the comparison of the experimental
and theoretical results for the total rates $\Gamma_{\text{in}}$ and $\Gamma_{\text{out}}$
as a function of $E_{A}$ for different values of the contact transmission.
One should remark the quite good agreement which is obtained for $\Gamma_{\text{out}}$
in the fast relaxation regime $(E_{A}>0.5 \Delta)$ and for $\Gamma_{\text{in}}$
in the slow relaxation regime. The drop in $\Gamma_{\text{out}}$ by more
than two orders of magnitude at $E_{A}\sim0.5\Delta$ is also captured
by our model. In the slow relaxation regime the model correctly describes
the decrease of $\Gamma_{\text{in}}$ and the increase in $\Gamma_{\text{out}}$
which is observed at fixed $E_{A}$ with increasing transmission.
The largest discrepancies between model and experiment
are found for $\Gamma_{\text{in}}$ when $E_{A}\gtrsim0.5\Delta$
and for $\Gamma_{\text{out}}$ when $E_{A}\lesssim0.5\Delta$. It should
be noticed, however, that the experimental determination of $\Gamma_{\text{in}}$ is less
precise for $E_A > 0.5\Delta$ where $P_{\infty} \ll 1$. The decay of $\Gamma_{\text{in}}$ in theory is essentially related to the assumption of a thermal distribution of quasiparticles, which leads to $\Gamma_{\text{in}}\propto \text{exp}(-(E_{A}+h \nu_{p}) / k_{B} T_{qp}).$ A weaker dependence on energy would be obtained with a self-consistent description of this distribution \cite{martinis2}.
On the other hand, the deviations found
for $\Gamma_{\text{out}}$ in the slow relaxation regime are more significant
and could indicate that some additional relaxation mechanism, like tunneling to vortices or quasiparticle traps in the vicinity of the contact, could be contributing for small $E_{A}$. 
Since in the limit where $\Gamma_{e^{*}\rightarrow e}\gg\Gamma_{\text{in}},\Gamma_{\text{out}}
\gg\Gamma_{e\rightarrow e^*}$,
the stationary occupation probability $P_{\infty}$ is simply given by $2\Gamma_{\text{in}}/(3\Gamma_{\text{in}}+\Gamma_{\text{out}})$ \cite{supplemental-longlived},
the deviations for $\Gamma_{\text{out}}$ explain why the theory overestimates $P_{\infty}$
in this regime. Further combined experimental and theoretical work
would be required to clarify this point.

\section{Conclusions}
\label{conclusions}

In conclusion, we have presented a theory which describes the dynamics of trapping and untrapping quasiparticles in phase-biased superconducting weak links. It is shown that in realistic conditions this dynamics can be controlled by the coupling of the weak link to its electromagnetic environment. The results are in semi-quantitative agreement with the experiments of Ref.~\onlinecite{longlived}, where the sharp jump observed in the trapping and untrapping rates is associated to the onset of the coupling to the environment plasma mode. The mechanisms described here can be relevant for controlling decoherence in superconducting qubits involving channels with non-negligible
transmission. In case of the Andreev qubits discussed in Refs. \onlinecite{marce,shumeiko}, where poisoning by trapped quasiparticles in the ABS should be avoided, the presence of
a mode of energy larger than $\Delta-E_A$ would be beneficial. In contrast, for the proposals of Refs. \onlinecite{nazarov,Padurariu} which are based on the manipulation of the odd states, a larger lifetime of the trapped quasiparticles is desirable. In this case one would need an electromagnetic environment containing no mode of frequency larger than $\Delta-E_A$. Finally, as discussed in Ref. \onlinecite{catelani}, even in the case of
qubits based on tunnel junctions changes in the occupation of the Andreev states make
the Josephson coupling and hence the qubit frequency fluctuate thus giving rise to 
dephasing. Therefore, even in this case of Andreev levels very close to the gap edge,
slowing down the dynamics of these occupations could have an influence on the qubit
decoherence. 

\begin{acknowledgements}
The authors acknowledge fruitful discussions with J. Martinis, R.
Egger, A. Zazunov, D. Urban, J. C. Cuevas, F. S. Bergeret and A. Mart\'{\i}n-Rodero. Financial
support by EU FP7 SE2ND project, Spanish Mineco project FIS2011-26516, ANR
contracts DOCFLUC and MASH, C\textquoteright{}Nano and by the People
Programme (Marie Curie Actions) of the European Union\textquoteright{}s
Seventh Framework Programme (FP7/2007-2013) under REA grant agreement
no. PIIF-GA-2011-298415 is acknowledged.
\end{acknowledgements}

\appendix

\section{Diagonalisation of the SC Hamiltonian and SC wavefunctions}
\label{appendixA}

The point contact is modelled as a 1D SNS junction with a Dirac delta
barrier in the normal region.

The normal region's length $\tilde{L}_N$ can be taken to the limit
$\tilde{L}_N\rightarrow 0$ in the ballistic regime, effectively turning the
scattering problem into the problem of a delta barrier in a superconducting
system, with a well-defined phase bias between the left and right leads.

{\it Continuum wavefuntions:} 
For an homogeneous superconducting system, the Hamiltonian's eigenfunctions
have a momentum $\left|k\right|=k_F\pm\kappa_{E}$, where
$\kappa_E=\xi_0^{-1}\sqrt{\left(E/\Delta\right)^2-1}$ ($\xi_0$ being the
superconducting coherence length, given by $hv_F/\Delta$ in the ballistic regime).
The eigenfunctions with positive energy $E_k$ and spin-up take the following shape:
\begin{equation}
\label{PlaneWaves}
\begin{array}{c}
\psi_{k,\tilde{\varphi}}^{\left(\uparrow\right)}\left(x\right)=\frac{1}{\sqrt{L}}
\left(\begin{array}{c} u^{}_{E}\\v^{}_{E} e^{i\tilde{\varphi}}\end{array}\right) e^{+ikx}
\begin{array}{c} \text{ for } \left|k\right|>k_F \\ \text{(quasielectrons)} \end{array}
\\ \\
\psi_{k,\tilde{\varphi}}^{\left(\uparrow\right)}\left(x\right)=\frac{1}{\sqrt{L}}
\left(\begin{array}{c} v^{}_{E}\\u^{}_{E} e^{i\tilde{\varphi}}\end{array}\right) e^{-ikx}
\begin{array}{c} \text{ for } \left|k\right|<k_F \\ \text{(quasiholes)} \end{array}
\end{array}
\end{equation}
\noindent where $\tilde{\varphi}$ is the phase of the superconducting order parameter,
$L$ is a length over which the freely propagating eigenfunctions are defined and
$u^{}_{E}$, $v^{}_{E}$ are, respectively, the electron and hole components of the plane wave:
\begin{equation}
\left(u/v\right)_{E}=\frac{1}{\sqrt{2}}\sqrt{1\pm\sqrt{1-\left(\Delta/E\right)^2}}
\end{equation}

\begin{figure}[ht!]
\includegraphics[scale=.50]{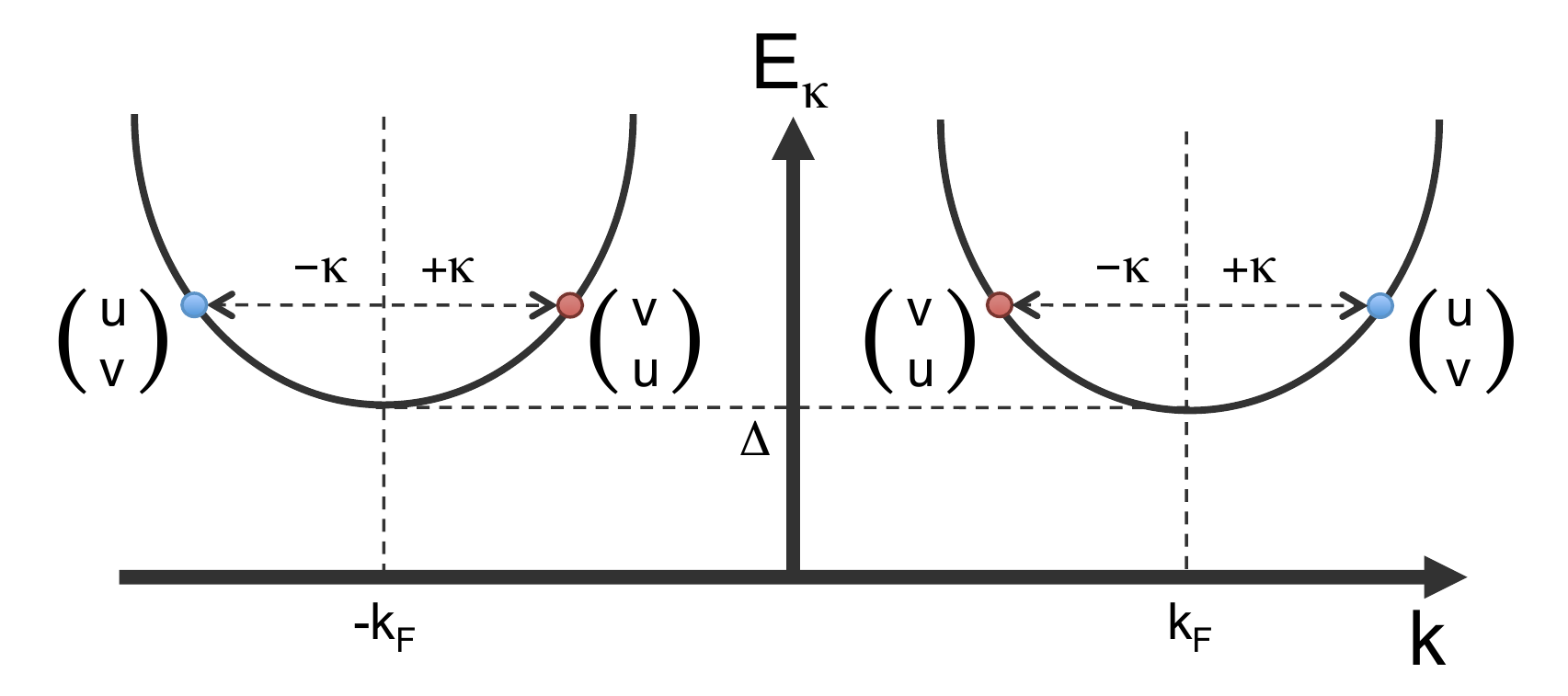}
\caption{Quasiparticle energies and two-component eigenfunctions near the Fermi level for spin-up
quasiparticles. There is a four-fold degeneracy of states for each energy $E$, with two quasielectron
states (blue dots) with a momentum $\left|k\right|=k_F+\kappa_E$ and another two quasiholes
with an absolute momentum of $\left|k\right|=k_F-\kappa_E$.}
\label{DispersionRelationship}
\end{figure}

\noindent These plane waves are schematically shown in 
Fig.~\ref{DispersionRelationship}.

Summations in momenta such as the one in Eq.~(\ref{golden-rule}) may be rewritten, for the sake
of convenience, as integrals over quasiparticle energies weighted
by the superconducting density of states
$\rho_{SC}\left(E\right)=\rho_F\frac{\left|E\right|}{\sqrt{E^2-\Delta^2}}$,
where $\rho_F$ is the normal density of states at the Fermi level.

That being the case we shall brand the wavefunctions (and the states they refer to) not by using
their momentum $k$ as an index, but their energy $E_k\rightarrow E$, their associated quasielectron/hole
character and the direction of their momentum.

With the plane waves from Eq. (\ref{PlaneWaves}) it is possible to construct
solutions to the BdeG equations in an inhomogeneous system following a scattering
approach \cite{BTK}.

These wavefunctions are Nambu spinors of the form
\begin{equation}
\label{GeneralWvF}
\psi_{E\uparrow}^{\left(\eta\pm\right)}\left(x\right)=
\left(\begin{array}{c} U_E^{\left(\eta\pm\right)}\left(x\right) \\
V_E^{\left(\eta\pm\right)}\left(x\right) \end{array} \right)
\end{equation}
\noindent where $U_E$ is the electron amplitude, $V_E$ the hole amplitude, and $\eta$ denotes
the electron/hole character of the quasiparticle state. Spin-down eigenfunctions can be easily
obtained from spin-up ones by use of the electron-hole symmetry in the system, through the substitution
$U\rightarrow V^*$, $V\rightarrow -U^*$.

For the sake of simplicity, we shall ommit the spin sub-index in the wavefunctions that we discuss next.

Condensing all the phase difference $\delta$ in the right lead, the wavefunctions
take the following shape

\begin{figure}[h!]
\includegraphics[scale=.45]{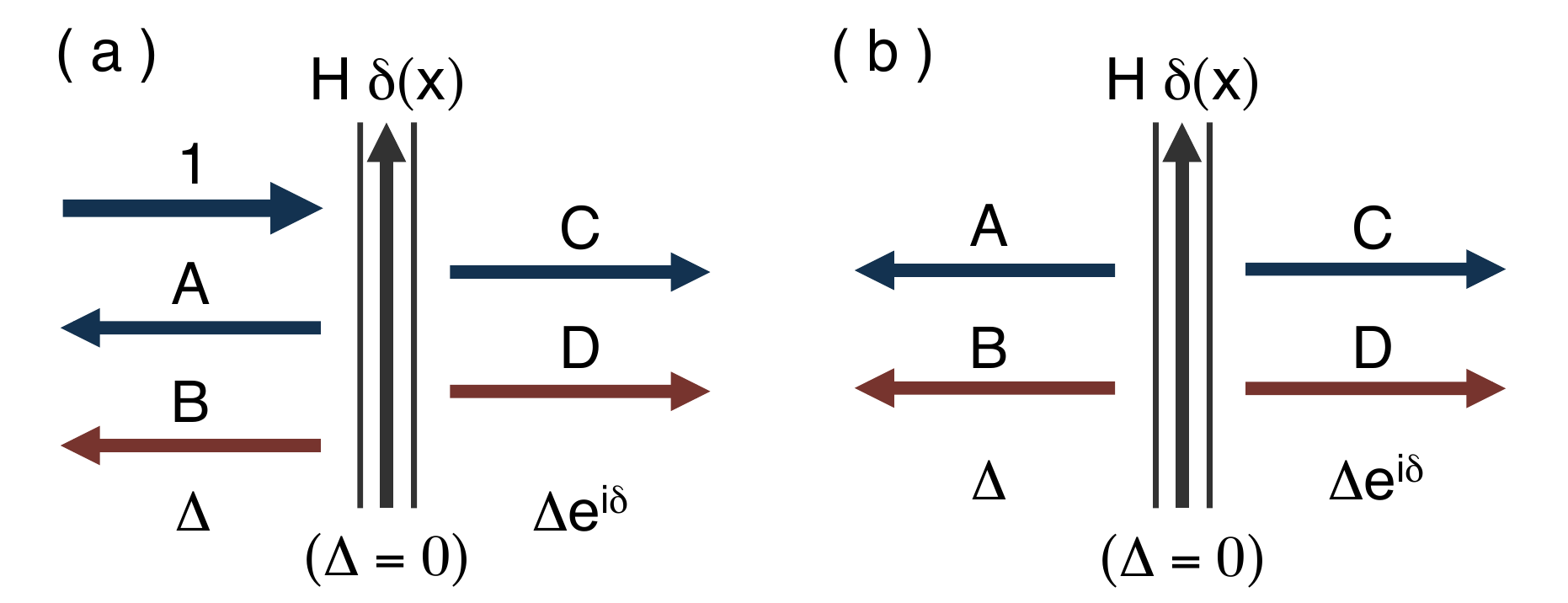}
\caption{Schematic representation of the scattering problem for the case of a
quasi-electron impinging the contact and the resulting outgoing quasiparticle
partial waves (a), either normally or Andreev-reflected or transmitted.
The scattering problem for the ABS does not require a source term (b),
but these outgoing partial waves exhibit an evanescent behaviour.}
\label{Scattering}
\end{figure}

\begin{widetext}
\begin{equation}
\psi_E^{\left(\eta\pm\right)}\left(x\right)=\psi_{Src}^{\left(\eta\pm\right)}\left(x\right)
+
\left(A_{\left(\eta\pm\right)}\psi_{E,0}^{\left(e-\right)}\left(x\right)
	 +B_{\left(\eta\pm\right)}\psi_{E,0}^{\left(h-\right)}\left(x\right)\right)
	 \Theta\left(-x\right)
+
\left(C_{\left(\eta\pm\right)}\psi_{E,\delta}^{\left(e+\right)}\left(x\right)
	 +D_{\left(\eta\pm\right)}\psi_{E,\delta}^{\left(h+\right)}\left(x\right)\right)
	 \Theta\left(x\right) \;,
\end{equation}
where $\psi_{Src}^{\left(\eta\pm\right)}$ is a source term of an quasielectron or 
quasihole
impinging the contact from any of the leads

\begin{equation}
\psi_{Src}^{\left(e+\right)}=\psi_{E,0}^{\left(e+\right)}\left(x\right)\Theta\left(-x\right)
\text{,}\hspace{5mm}
\psi_{Src}^{\left(h+\right)}=\psi_{E,0}^{\left(h+\right)}\left(x\right)\Theta\left(-x\right)
\text{,}\hspace{5mm}
\psi_{Src}^{\left(e-\right)}=\psi_{E,\delta}^{\left(e-\right)}\left(x\right)\Theta\left(x\right)
\text{,}\hspace{5mm}
\psi_{Src}^{\left(h-\right)}=\psi_{E,\delta}^{\left(h-\right)}\left(x\right)\Theta\left(x\right) \;.
\end{equation}

The rest of the contributions to the wavefunctions are outgoing partial waves
(as illustrated in Fig.~\ref{Scattering}).

Imposing continuity for the wavefunction and its derivative (taking into account the effect
of the delta barrier) the values of the partial wave coefficients are obtained
for each incidence

\begin{equation}
\begin{array}{cc}
A_{\left(e+\right)}=-\mu_\tau\tau\left(\mu_\tau+i\right)\sinh^2\theta_E\cdot Q^{-1} & 
B_{\left(e+\right)}=i\tau\sin\delta/2\sinh\left(\theta_E+i\frac{\delta}{2}\right)\cdot Q^{-1} \\ \\
C_{\left(e+\right)}=-i\tau\left(\mu_\tau+i\right)e^{-i\delta/2}
					\sinh\theta_E\sinh\left(\theta_E+i\frac{\delta}{2}\right)\cdot Q^{-1} &
D_{\left(e+\right)}=-\mu_\tau\tau e^{-i\delta/2}\sinh\theta_E\sin\frac{\delta}{2}\cdot Q^{-1}
\end{array}
\end{equation}
\begin{equation}
\begin{array}{c}
A_{\left(e-\right)}=C_{\left(e+\right)}\left(-\delta\right) \\\\
B_{\left(e-\right)}=D_{\left(e+\right)}\left(-\delta\right) \\\\
C_{\left(e-\right)}=A_{\left(e+\right)}\left(-\delta\right) \\\\
D_{\left(e-\right)}=B_{\left(e+\right)}\left(-\delta\right)
\end{array}
\hspace{7.5mm}
\begin{array}{c}
A_{\left(h+\right)}=B_{\left(e+\right)}^* \\\\
B_{\left(h+\right)}=A_{\left(e+\right)}^* \\\\
C_{\left(h+\right)}=D_{\left(e+\right)}^* \\\\
D_{\left(h+\right)}=C_{\left(e+\right)}^*
\end{array}
\hspace{7.5mm}
\begin{array}{c}
A_{\left(h-\right)}=C_{\left(h+\right)}\left(-\delta\right) \\\\
B_{\left(h-\right)}=D_{\left(h+\right)}\left(-\delta\right) \\\\
C_{\left(h-\right)}=A_{\left(h+\right)}\left(-\delta\right) \\\\
D_{\left(h-\right)}=B_{\left(h+\right)}\left(-\delta\right)
\end{array} \;,
\end{equation}
\end{widetext}
where $\sinh\theta_E=\sqrt{(\frac{E}{\Delta})^2-1}$, $\mu_\tau=\sqrt{\frac{R}{\tau}}$,
$\tau$ is the normal transmission probability from the potential barrier,
$R=1-\tau$ its normal reflection probability, and $Q=\sinh^2\theta_E+\tau\sin^2\frac{\delta}{2}$.

{\it Andreev Bound State wavefunction:}
The wave amplitudes for states with $E<\Delta$ exhibit subgap poles
at $E=E_A\left(\delta\right)=\Delta\sqrt{1-\tau\sin^2\frac{\delta}{2}}$,
which signals the existence of a bound state at such an energy.

The wavefunctions for the Andreev Bound States (ABS) may be obtained in a
similar way than those for states lying at $E>\Delta$ taking into account
that the quasiparticle momentum gains an imaginary component below the gap.
The partial waves for the ABS are
\begin{equation}
\label{PlaneWavesABS}
\begin{array}{cc}
\psi_{E_A,\tilde{\varphi}}^{\left(e\pm\right)}\left(x\right)=\frac{1}{\sqrt{\xi_0}}
\left(\begin{array}{c} u_A\\v_A e^{i\tilde{\varphi}}\end{array}\right)
e^{\pm i\left(k_F+i\kappa_A\right)x} \\ \\
\psi_{E_A,\tilde{\varphi}}^{\left(h\pm\right)}\left(x\right)=\frac{1}{\sqrt{\xi_0}}
\left(\begin{array}{c} v_A\\u_A e^{i\tilde{\varphi}}\end{array}\right)
e^{\mp i\left(k_F-i\kappa_A\right)x}
\end{array}
\end{equation}
These differ from the propagating partial waves because the $u_E$, $v_E$
coefficients and the quasiparticle momentum $\kappa_E$ become complex for
$E<\Delta$:
\begin{gather}
u_E\rightarrow u_A=\frac{1}{\sqrt{2}}e^{i\theta_A/2}
\hspace{5mm}
v_E\rightarrow v_A=\frac{1}{\sqrt{2}}e^{-i\theta_A/2}
\\
\kappa_E\rightarrow i\kappa_A = i\xi_0^{-1}\sin\theta_A
\\
\text{with }\sin\theta_A=\sqrt{\tau}\left|\sin\frac{\delta}{2}\right|=
\frac{\sqrt{\Delta^2-E_A^2}}{\Delta}\text{.}
\end{gather}

Only partial waves confined within a length $\kappa_A^{-1}$,
which diverges for $E_A\rightarrow\Delta$, may appear in the wavefunctions
\begin{equation}
\psi_A\left(x\right)=
\begin{array}{c}
\left(A_A\psi_{E_A,0}^{\left(e-\right)}\left(x\right)
	 +B_A\psi_{E_A,0}^{\left(h-\right)}\left(x\right)\right)
  \Theta\left(-x\right) \\\\
+ \left(C_A\psi_{E_A,\delta}^{\left(e+\right)}\left(x\right)
	   +D_A\psi_{E_A,\delta}^{\left(h+\right)}\left(x\right)\right)
  \Theta\left(x\right)
\end{array}\text{.}
\end{equation}

A linear homogeneous system of equations is obtained for the partial
wave weights by applying the same conditions as in the case of the
continuum states. The system exhibits a nontrivial solution for
$E=E_A\left(\delta\right)$. Eliminating the redundant equation and
imposing the normalisation condition for the wavefunction, it is finally
obtained that, in the bound states
\begin{equation}
\begin{cases}
A_A=-i\tilde{N}_A\sin\left(\theta_A-\frac{\delta}{2}\right)e^{-i\tilde{\beta}_\tau}e^{i\delta/2}
\\\\
B_A=-i\sqrt{R}\tilde{N}_A\sin\frac{\delta}{2}\cdot e^{i\tilde{\beta}_\tau}e^{i\delta/2}
\\\\
C_A=\sqrt{R}\tilde{N}_A\left|\sin\frac{\delta}{2}\right|e^{-i\tilde{\beta}_\tau}
\\\\
D_A=\sigma_\delta\tilde{N}_A\sin\left(\theta_A-\frac{\delta}{2}\right)e^{i\tilde{\beta}_\tau}
\end{cases}
\end{equation}
where $\sin\tilde{\beta}_\tau=\sqrt{\tau}$, $\sigma_\delta=\text{sign}\left(\delta\right)$ and
\begin{equation}
\tilde{N}_A = 
\sqrt{\frac{-\sigma_\delta\sqrt{\tau}}{2\cos\theta_A\sin\left(\theta_A-\frac{\delta}{2}\right)}}
\text{.}
\end{equation}

A similar derivation of this result can be found in \onlinecite{landrythesis}.

\begin{figure}
\includegraphics[scale=.50]{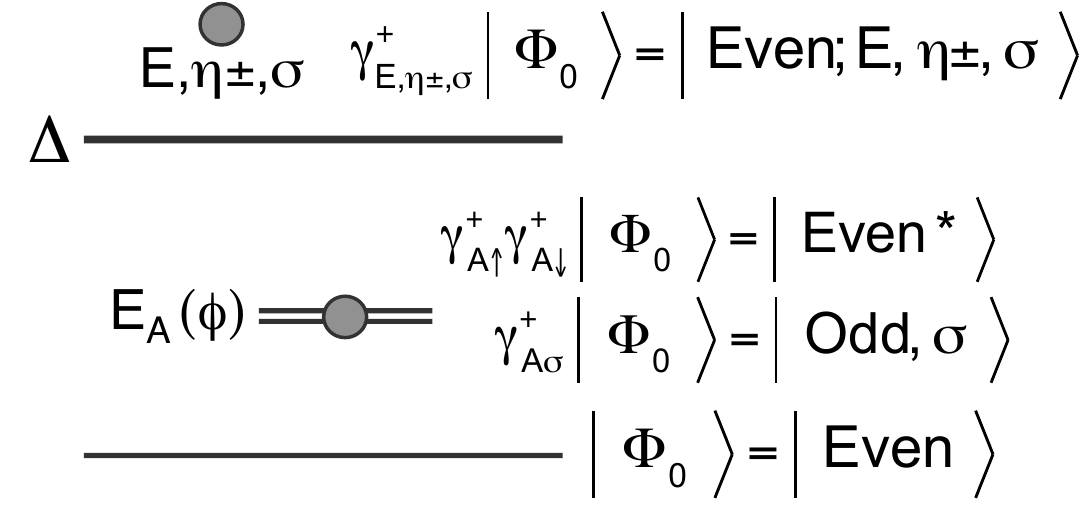}
\caption{Single-particle excitations of the unperturbed SC Hamiltonian,
with the parity notation for different ABS occupancies.
The ground state of the system is noted as the even state. As single-particle
excitations of the ABS have an odd parity, the ABS doublet excitation has an even parity,
so it is branded as the even excited $\left|\text{Even}^*\right\rangle$ state.}
\label{FigExcitts}
\end{figure}

The solutions of the BdeG equations allow us to express the electron field operators as

\begin{widetext}
\begin{equation}
\label{B-V_transformation}
\Psi_\sigma\left(x\right)=U_A\left(x\right)\gamma^{}_{A\sigma}
				  -\sigma V^*_A\left(x\right)\gamma^\dagger_{A\bar{\sigma}}
+\sum\limits_{E,\eta\pm}\left(U_E^{\left(\eta\pm\right)}\left(x\right)\gamma^{}_{E,\eta\pm,\sigma}
-\sigma V^{\left(\eta\pm\right)*}_E\left(x\right)\gamma^\dagger_{E,\eta\mp,\bar{\sigma}}\right)
\end{equation}
where $\gamma_{A\sigma}$ and $\gamma_{E,\eta\pm,\sigma}$ are the quasiparticle operators
which diagonalise the SC Hamiltonian. The excitation spectrum of the system is represented
in Fig.~\ref{FigExcitts}.

\section{Current operator and transition rates}
\label{appendixB}
The current operator in the new basis defined by Eq. (\ref{B-V_transformation}) is
\begin{equation}
\label{CurrentOperator}
\hat{I}\left(x\right)=-\frac{\hbar e}{2mi}\sum_{i,j,\sigma}
\left(\begin{array}{cc}\gamma^\dagger_{i\sigma} & \gamma^{}_{i\bar{\sigma}}\end{array}\right)
\left(\begin{array}{cc}
U^*_i\frac{dU_j}{dx}-\frac{dU^*_i}{dx}U_j &
\sigma\left(U^*_i\frac{dV^*_{-j}}{dx}-\frac{dU^*_i}{dx}V^*_{-j}\right)
\\
\sigma\left(V_{-i}\frac{dU_j}{dx}-\frac{dV_{-i}}{dx}U_j\right) &
V_{-i}\frac{dV^*_{-j}}{dx}-\frac{dV_{-i}}{dx}V^*_{-j}
\end{array}\right)
\left(\begin{array}{c}\gamma^{}_{j\sigma} \\ \gamma^\dagger_{j\bar{\sigma}}\end{array}\right)
\end{equation}
The sum in $i$, $j$ indices are a shorthand notation for all the different contributions
appearing in Eq. (\ref{B-V_transformation}). The minus sign in front of some particular
wavefunction subindices notes that such a component corresponds to the antiparall wavefunction
(e.g., if $i=E\left(e+\right)$, then $U_{-i}=U_E^{\left(e-\right)}$), which only applies when
the index corresponds to an excitation in the continuum.
\end{widetext}

\subsection{Transitions involving the odd states}

A relevant matrix element in the problem is the one associated to the
$\left|\text{Odd},\sigma\right\rangle \rightarrow \left|\text{Even};E,e+,\sigma\right\rangle$
process, which is found to be
\begin{equation}
-\frac{\hbar ek_F}{m\sqrt{L\xi_0}}
\left(C_AC^*_{\left(e+\right)}-D_AD^*_{\left(e+\right)}\right)
\left(u_Eu_A+v_Ev_A\right)\text{.}
\end{equation}
The electron-hole symmetry in the field transformations ensures that this matrix
element is the complex conjugate of the matrix element associated to the process
$\left|\text{Odd},\sigma\right\rangle \rightarrow \left|\text{Even};E,h-,\sigma\right\rangle$.

In the ballistic limit $\tau\rightarrow 1$, restricting $\delta$ to the interval
$\left[0,\pi\right]$ so as to establish a bijection
between $\delta$ and $E_A\left(\delta\right)$, one finds
\begin{gather}
\left|C_A\right|\rightarrow\left(\frac{\Delta^2-E_A^2}{\Delta^2}\right)^{1/4}
\hspace{5mm}
D_A\rightarrow 0
\\
\left|C_{\left(e+\right)}\right|\rightarrow\sqrt{\frac{E^2-\Delta^2}{E^2-E_A^2}}
\hspace{5mm}
D_{\left(e+\right)}\rightarrow 0\text{.}
\end{gather}
The squared amplitude of these matrix elements in this limit is
\begin{equation}
\label{MatrixElem1}
\frac{\hbar^2e^2k_F^2}{m^2L\xi_0}
\frac{\sqrt{\Delta^2-E_A^2}}{\Delta}
\frac{E^2-\Delta^2}{E^2-E_A^2}
\left(1+\frac{E_A}{E}\right) \text{.}
\end{equation}
Matrix elements for the
$\left|\text{Odd},\sigma\right\rangle \rightarrow \left|\text{Even};E,e-,\sigma\right\rangle$
and $\left|\text{Odd},\sigma\right\rangle \rightarrow \left|\text{Even};E,h+,\sigma\right\rangle$
processes vanish in the limit of perfect transmission.

Conversely, the squared amplitude for the quasiparticle recombination processes
$\left|\text{Odd},\sigma;E,e-,\bar{\sigma}\right\rangle \rightarrow \left|\text{Even}\right\rangle$
and $\left|\text{Odd},\sigma;E,h+,\bar{\sigma}\right\rangle \rightarrow \left|\text{Even}\right\rangle$
in the limit $\tau\rightarrow 1$ is found to be
\begin{equation}
\label{MatrixElem2}
\frac{\hbar^2e^2k_F^2}{m^2L\xi_0}  \frac{\sqrt{\Delta^2-E_A^2}}{\Delta}
\frac{E^2-\Delta^2}{E^2-E_A^2}     \left(1-\frac{E_A}{E}\right) \text{.}
\end{equation}

\noindent Whereas the amplitudes for the other two recombination processes, which are
$\left|\text{Odd},\sigma;E,e+,\bar{\sigma}\right\rangle \rightarrow \left|\text{Even}\right\rangle$
and $\left|\text{Odd},\sigma;E,h-,\bar{\sigma}\right\rangle \rightarrow \left|\text{Even}\right\rangle$,
are zero in this same limit.

The products of the terms that contain in these expressions the functional
dependence in $E$ and $E_A$ with the superconducting density of states yield
the factors $g\left(E,E_A\right)$ mentioned in the main article
\begin{equation}
g \left( E, E_A \right) =
\frac{\sqrt{\left( E^2-\Delta^2\right)\left(\Delta^2-E_A^2\right)}}
{ \Delta \left( E - E_A \right) }\text{.}
\end{equation}

In the opposite tunnel limit $\tau\rightarrow 0$ all squared amplitudes tend to zero
as $\tau^{3/2}$, with the leading term being the same for the four different processes
\begin{equation}
\frac{\hbar^2e^2k_F^2}{m^2L\xi_0}
\frac{\tau}{2}
\frac{\sqrt{\Delta^2-E_A^2}}{\Delta}
\left(1 + \frac{\Delta\cos\delta}{E}\right) \text{.}
\end{equation}

The same applies to amplitudes of recombination processes, with a minus sign appearing
inside the parenthesis instead of a plus.

From these we may define another $g\left(E,E_A\right)$ factor for the tunnel regime.
\begin{equation}
g \left( E,E_A \right) = \frac{\tau}{2}
\sqrt{\frac{\Delta^2-E_A^2}{E^2-\Delta^2}}
\left(\frac{E}{\Delta}+\text{sgn}\left(E_A\right)\cos\delta\right)\text{.}
\end{equation}

These results are in agreement with the ones recently derived by Kos et al. using a different method
\cite{KosGlazman} taking also into account the factor $\left(E\pm E_A\right)^{-1}$ that comes from
the environmental density of states (see Eq. (\ref{EMDOS})).

\subsection{Transitions between the even states}

An analytical expression for the amplitude that links the two even
states can be derived for any value of $\tau$
\begin{eqnarray}	
\left\langle\text{Even}\left|\hat{I}\right|\text{Even}^*\right\rangle=
\frac{\hbar ek_F}{m\xi_0}C_AD_A\left(u_A^2-v_A^2\right)e^{i\delta}
\\ \nonumber \\
\left|\left\langle\text{Even}\left|\hat{I}\right|\text{Even}^*\right\rangle\right|^2=
\frac{e^2\Delta^4}{\hbar^2}\frac{\left(1-\tau\right)\tau^2\sin^4\frac{\delta}{2}}
{E_A^2\left(\delta\right)} \;.
\end{eqnarray}
These results coincide with the results from \onlinecite{shumeiko}.

\section{Modelling the EM environment for the experiment in Ref. [1]}
\label{appendixC}

\begin{figure}
\includegraphics[scale=0.50]{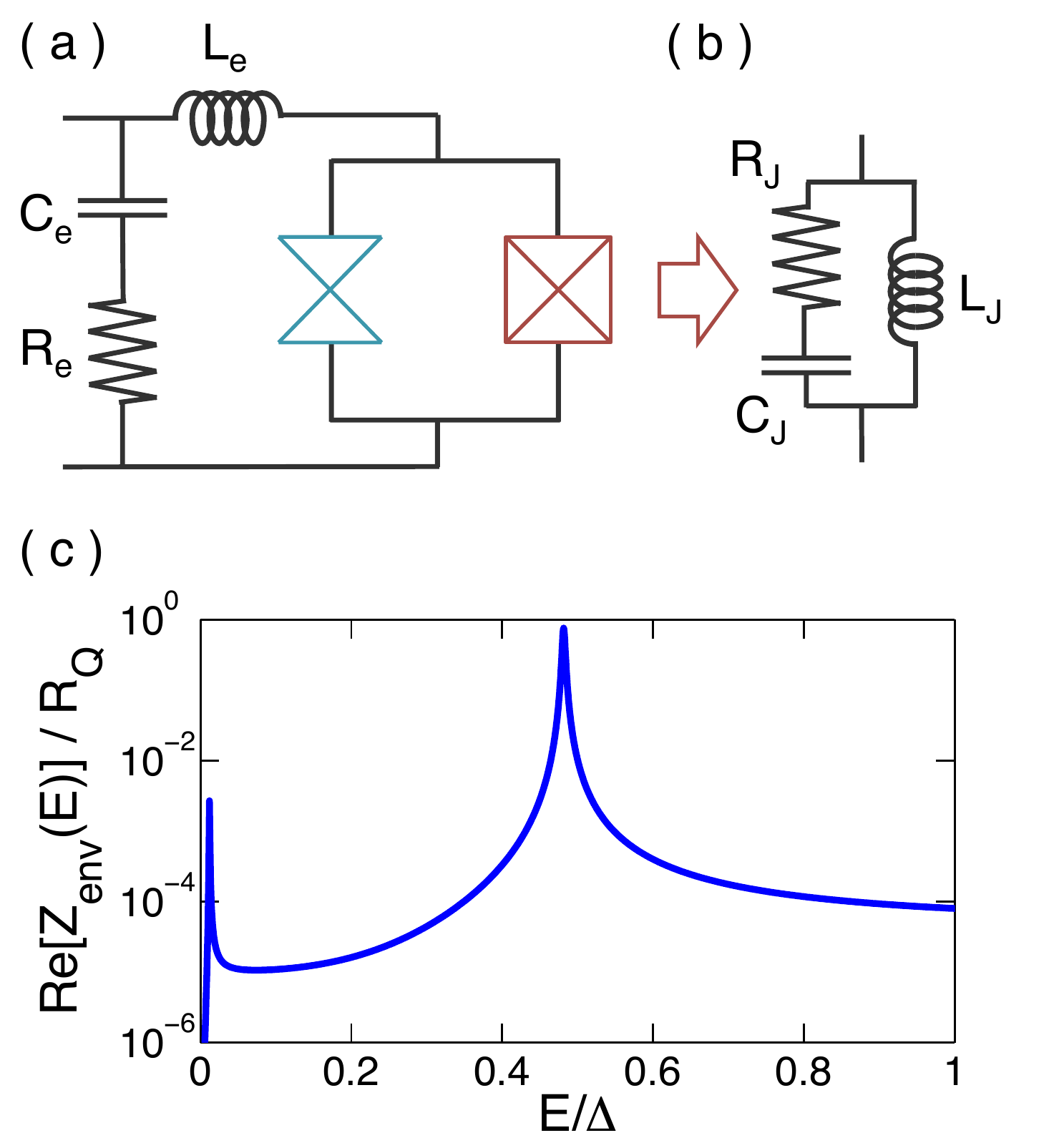}
\caption{(a) Electrical elements in the relevant neighborhood of
the SQUID loop. (b) Equivalent electrical model for the Josephson 
junction in the SQUID. (c) Environmental impedance as seen by the SC.}
\label{EMDoSCircuit}
\end{figure}

The density of environmental modes $D\left(h\nu\right)$ is, following
the formalism presented in Ref.~\onlinecite{IngoldNazarov},
\begin{equation}
D\left(h\nu\right)=\frac{1}{h\nu}
\frac{\text{Re}\left\lbrace Z_{env}\left(\nu\right)\right\rbrace}{R_Q}\text{,}
\label{EMDOS}
\end{equation}
where $Z_{env}\left(\nu\right)$ is the electric impedance as seen from the SC,
represented in Fig.~\ref{EMDoSCircuit}(c).

In the experiment shown in Ref.~\onlinecite{longlived}, a SC is placed in parallel with
a Josephson junction having a critical current much larger than that of the atomic contact
(see Fig.~\ref{EMDoSCircuit}(b)). This junction is perceived by
the atomic contact as the parallel combination of an inductor $L_J=\varphi_0/I_0$ and a
capacitor $C_J$, $I_0$ being the critical current of the junction.
The finite quality factor of the corresponding electromagnetic (``plasma")
mode is modeled with a resistance $R_J$ in series with the capacitor.
The SQUID loop formed by the contact and the junction is connected to a biasing circuit
through an inductor $L_e$ and a capacitance $C_e$. Dissipation in this circuit
is modelled by a resistance $R_e$ in series with the capacitor
(see Fig.~\ref{EMDoSCircuit}(a)).

The total impedance seen by the SC is
\begin{equation}
\frac{Z_{env}\left(\nu\right)}{R_Q}=\frac{4}{\pi}\frac{E_C}{h\nu_{p0}}
\frac{i \> \frac{\nu}{\nu_{p0}} a_e\left(\nu\right) b_{p0}\left(\nu\right)}
     { a_{p0}\left(\nu\right) a_e\left(\nu\right)
     - \frac{\nu^2}{\nu_3^2} \> b_{p0}\left(\nu\right)} \; ,
\end{equation}
with $ a_{\star} \left(\nu\right) = 1 + i \> \frac{1}{Q_{\star}} \> \frac{\nu}{\nu_{\star}}
                                - \frac{\nu^2}{\nu^2_{\star}} $ where $\star \equiv e, p0$ and
$ b_{p0} \left(\nu\right) = 1 + i \> \frac{1}{Q_{p0}} \> \frac{\nu}{\nu_{p0}} $.

The characteristic frequencies and quality factors inside these expressions are
\begin{gather}
\begin{array}{c}
\nu_{p0}=\frac{1}{2\pi}\left(L_JC_J\right)^{-1/2} \hspace{4mm}
\nu_e=\frac{1}{2\pi}\left(L_eC_e\right)^{-1/2}
\\\\
\nu_3=\frac{1}{2\pi}\left(L_JC_e\right)^{-1/2}
\end{array}
\\\nonumber
\\
Q_{p0}=\frac{1}{R_J}\sqrt{\frac{L_J}{C_J}} \hspace{4mm}
Q_e=\frac{1}{R_e}\sqrt{\frac{L_e}{C_e}} \text{.}
\end{gather}

The equivalent circuit in Fig.~\ref{EMDoSCircuit} has two modes. A low-frequency mode
determined essentially by the on-chip LC filter connecting the SQUID to the outside world,
and a high-frequency mode corresponding to the plasma oscillation of the junction ``dressed"
by the external circuit:
\begin{equation}
\nu_P=\frac{1}{2\pi}\sqrt{\frac{L_J^{-1}+L_e^{-1}}{C_J}}
\end{equation}

Parameters for this equivalent circuit were obtained in the following way:

The Josephson junction inductance $L_J=595$ pH is determined by the critical current extracted
from the switching probability measurements. The other five parameters of the equivalent
circuit of Fig.~\ref{EMDoSCircuit} were adjusted so as to reproduce at best
all the available experimental information.

\begin{enumerate}
\item{The energy gap measured from the IV characteristics is $\Delta=194$ $\mu$eV.}
\item{The dressed plasma frequency must be close to $\Delta/2h$ in order to explain
the position of the sharp threshold observed in the rates $\Gamma_{in}$ and $\Gamma_{out}$.
This is compatible with what was reported in \cite{nature-landry}.}
\item{A value of $C_e=60$ pF is expected from a measurement at very low frequency
($100$ kHz) on a larger test capacitor fabricated on the same run.}
\item{The DC sub-gap current of the JJ alone is $I_J=4.4$ nA at $V_J=\Delta/4$e.
Imposing the power equality $I_JV_J=R_JI_0$ between the DC injected power and the microwave
power absorbed at $\Delta/2h$ by the junction's environment, we get $R_J=0.3$ $\Omega$.}
\item{The low frequency mode of the environment was measured at $558$ MHz in a separate
microwave reflectometry experiment \cite{landrythesis}.}
\item{At this resonance frequency, the reflection amplitude $S_{11}$ shows a dip of $-15$ dB,
from which we determine $R_e=0.25$ $\Omega$.}
\end{enumerate}

The two capacitances, $C_J$ and $C_e$, and the environmental inductance $L_e$ were adjusted
so as to reproduce the two characteristic frequencies of the circuit. The chosen value
$C_J=168$ fF is a $25\%$ lower than what is expected from the nominal area of the junction
($2.8$ $\mu$m$^2$) and the typical specific capacitance for the junctions fabricated usually
in our laboratory ($75$ fF/$\mu$m$^2$). $C_e=68$ pF is a $13\%$ higher than
what is expected from the test low frequency measurement. Finally, the nominal value $L_e=600$ pH 
for the environmental inductance is an $80\%$ of what is expected from a crude geometrical estimation.

With these values we predict a dressed plasma mode frequency of $0.48\Delta/h$
and a quality factor $Q=116$.

\section{Relaxation due to phonons}
\label{appendixD}

The electron-phonon interaction in real space is \cite{FetterWalecka}
\begin{equation}
\hat{H}_\text{e-ph}=\tilde{\gamma}\int d\mathbf{r}\sum_\sigma
\Psi^\dagger_\sigma\left(\mathbf{r}\right)\Psi_\sigma\left(\mathbf{r}\right)
\hat{\phi}\left(\mathbf{r}\right)
\end{equation}
\noindent where $\hat{\phi}\left(\mathbf{r}\right)$ is the phonon field operator:
\begin{equation}
\hat{\phi}\left(\mathbf{r}\right)=\sum_{\mathbf{q}}\sqrt{\frac{h\nu_{\mathbf{q}}}{2V}}
\left(b^{}_{\mathbf{q}}e^{i\mathbf{qr}}+b^\dagger_{\mathbf{q}}e^{-i\mathbf{qr}}\right)\text{.}
\end{equation}

The electron-phonon coupling constant $\tilde{\gamma}$ is
$\frac{Z\hbar^2\pi^2}{mk_F}\frac{n_0}{B^{\frac{1}{2}}}$, with $n_0$ being the atomic density,
$B$ the adiabatic bulk modulus and $Z$ the electron valence from the superconductor.

The SC density inside the interaction Hamiltonian takes a form similar to the current operator's
in the quasiparticle basis (Eq. (\ref{CurrentOperator})). But differently from the
coupling with the EM environment, the phonon coupling depends on the geometrical spread of the SC
wavefunctions, a feature characteristic of the coupling of phonons with localized states
\cite{zazunov2005,IvanovFeigelman}.

After eliminating terms linear in $e^{\pm 2ik_Fx}$, which vanish in the spatial integration due
to their rapid oscillatory behavior, the matrix element associated to the process
$\left|\text{Odd},\sigma\right\rangle \rightarrow \left|\text{Even};E,e+,\sigma\right\rangle$
is found to be
\begin{equation}
\left(C^*_{\left(e+\right)}C_A-D^*_{\left(e+\right)}D_A\right)\left(u_Eu_A-v_Ev_A\right)
\frac{e^{-\left(\kappa_A+i\kappa_E\right)x}}{\sqrt{L\xi_0}}\text{.} \nonumber
\end{equation}
In the limit $\tau\rightarrow 1$, the squared amplitude of the part of this matrix element
that does not depend on $x$, as well as the similar quantity obtained from the matrix element
for the process
$\left|\text{Odd},\sigma\right\rangle \rightarrow \left|\text{Even};E,h-,\sigma\right\rangle$,
tend to the expression in Eq. (\ref{MatrixElem2}), except for the factor
$\left(\hbar ek_F/m\right)^2$.

It can also be found in the same limit that the analogous $x$-independent quantity for the
$\left|\text{Odd},\sigma;E,e-,\bar{\sigma}\right\rangle \rightarrow \left|\text{Even}\right\rangle$
and $\left|\text{Odd},\sigma;E,h+,\bar{\sigma}\right\rangle \rightarrow \left|\text{Even}\right\rangle$
processes is, on the other hand, the same as in Eq. (\ref{MatrixElem1}).

The spatial integrals are of the form
\begin{eqnarray}
\int_0^\infty dx\cdot e^{-(\kappa_A\pm i\kappa_E)x}\sin q_xx
\cdot F\left(\mathbf{q_\perp},x\right) \;,
\nonumber
\end{eqnarray}
where
\begin{eqnarray}
F\left(\mathbf{q_\perp},x\right)=\int_{A_\perp\left(x\right)} d^2\mathbf{r_\perp}
e^{i\mathbf{q_\perp r_\perp}}\left|\Psi_\perp\left(\mathbf{r_\perp}\right)\right|^2 \;,
\nonumber
\end{eqnarray}
$\Psi_\perp\left(\mathbf{r_\perp}\right)$ being the axial spread
of the SC wavefunctions on the leads, whose geometric details are
enclosed in their cross section $A_\perp\left(x\right)$.

The momentum transfer $Q$ to or from a phonon taking an active role
in these relaxation processes is large compared to the inverse
penetration length of the ABS: $Q\gg\kappa_A,\kappa_E$ in
the region $E_A\left(\delta\right)<\frac{\Delta}{2}$.
Following the approximations detailed in \onlinecite{zazunov2005}
in the theoretical description of the phonon-mediated
$\left|\text{Even}^*\right\rangle\rightarrow\left|\text{Even}\right\rangle$ relaxation,
the $F\left(\mathbf{q_\perp},x\right)$ factor
introduces a cutoff $\tilde{L}$ in the integral in the $x$ direction
\begin{equation}
F\left(\mathbf{q_\perp},x\right)\rightarrow\theta\left(\tilde{L}-|x|\right)
\end{equation}

\noindent and the spatial integration may be easily evaluated in the limit
$\tilde{L}^{-1}\gg Q\gg\kappa_A,\kappa_E$:
\begin{equation}
\frac{4Q\sin Q\tilde{L}}{Q^2+\left|\kappa_A\pm i\kappa_E\right|^2}\approx 4\tilde{L}\text{.} \nonumber
\end{equation}

Combining the different contributions, the squared amplitude for the process
$\left|\text{Odd},\sigma\right\rangle \rightarrow \left|\text{Even};E,e/h\pm,\sigma\right\rangle$
mediated by the emission of a phonon is

\begin{equation}
\left|M_1\right|^2 = 8\frac{h\nu_Q}{V}\frac{\tilde{L}^2\tilde{\gamma}^2}{L\xi_0}
\sqrt{1-\frac{E_A}{\Delta}^2}\frac{E^2-\Delta^2}{E^2-E_A^2}\frac{E-E_A}{E}\text{.} \nonumber
\end{equation}

The transition rate for such a process is
\begin{widetext}

\begin{equation}
\frac{2\pi}{\hbar} \> V
\int\frac{d^3\mathbf{Q}}{\left(2\pi\right)^3} \> 4\left|M_1\right|^2
\rho_{SC}\left(E\right) \left(1-f_{\text{FD}}\left(E,T_{\text{qp}}\right)\right)
\> f_{\text{BE}}\left(h\nu_\mathbf{Q},T_{\text{ph}}\right) \> \delta\left(h\nu_\mathbf{Q}-E+E_A\right)\text{,}
\label{PhononRate1}
\end{equation}
where we have used the same notation as in the main text.
We may rewrite the integral over momenta in Eq. (\ref{PhononRate1})
as an integral over energies, with a density of states quadratic in $E-E_A$
that appears as a result of this transformation. The resulting total rate is then
\begin{equation}
\Gamma_{\text out}^{\left(\text{a}\right)} = \frac{2\pi}{\hbar} \frac{8}{\pi}
\left(\frac{\tilde{L}}{\xi_0}\right)^2 \frac{\tilde{\gamma}^2}{\pi^2}
\left(\frac{\Delta}{\hbar c_s}\right)^3 \int_\Delta^\infty \frac{dE}{\Delta}
\left(\frac{E-E_A}{\Delta}\right)^3 g\left(E,-E_A\right)
\> f_{\text{BE}}\left(E-E_A,T_{\text{ph}}\right) \> \left(1-f_{\text{FD}}\left( E,
T_{\text {qp}}\right)\right)
\text{.}
\end{equation}

Repeating the same process for the process for the form
$\left|\text{Odd},\sigma;E,\eta\bar{\sigma}\right\rangle \rightarrow \left|\text{Even}\right\rangle$
yields

\begin{equation}
\Gamma_{\text out}^{\left(\text{b}\right)} = \frac{2\pi}{\hbar} \frac{8}{\pi}
\left(\frac{\tilde{L}}{\xi_0}\right)^2 \frac{\tilde{\gamma}^2}{\pi^2}
\left(\frac{\Delta}{\hbar c_s}\right)^3 \int_\Delta^\infty \frac{dE}{\Delta}
\left(\frac{E+E_A}{\Delta}\right)^3 g\left(E,E_A\right) \>
\left(1-f_{\text{BE}}\left(E+E_A,T_{\text{ph}}\right)\right) \> f_{\text{FD}}\left(E,
T_{\text{qp}}\right)
\text{.}
\end{equation}
\end{widetext}

Using the material constants for aluminum, the factor
$\frac{2\pi}{\hbar}\frac{\tilde{\gamma}^2}{\pi^2}\left(\frac{\Delta}{\hbar c_s}\right)^3$
is of the order of $10$ GHz. The phonon rate is reduced because of the relevant thermal factors
(which are of the order of $e^{-\beta\Delta}\sim 10^{-5}$) and the geometrical factor
$\frac{8}{\pi}\left(\frac{\tilde{L}}{\xi_0}\right)^2$. If this factor is of the order of $10^{-2}$,
the phonon-induced relaxation rates are reduced to around 1~kHz, which coincide with other
estimations in analogous systems \cite{Padurariu}.

%----------------------------------------------------------------------------------------


\begin{thebibliography}{10}
\bibitem{comment} Similar processes but for photons from an external driving field
were considered in Ref.~\onlinecite{spectro} based on the results of 
Ref.~\onlinecite{KosGlazman}.
\end{thebibliography}
\end{document}